# Source anisotropies and pulsar timing arrays

Bruce Allen[*]

Max Planck Institute for Gravitational Physics (Albert Einstein Institute), Leibniz Universität Hannover,
Callinstrasse 38, D-30167, Hannover, Germany

Deepali Agarwal[†]

Centre for Cosmology, Particle Physics and Phenomenology (CP3), Université Catholique de Louvain,
Louvain-la-Neuve, B-1348, Belgium

Joseph D. Romano[‡]

Department of Physics and Astronomy, University of Texas Rio Grande Valley,
One West University Boulevard, Brownsville, Texas 78520, USA

Serena Valtolina[§]

Max Planck Institute for Gravitational Physics (Albert Einstein Institute), Leibniz Universität Hannover,
Callinstrasse 38, D-30167, Hannover, Germany

Pulsar timing arrays (PTAs) hunt for gravitational waves (GWs) by searching for the correlations that GWs induce in the time-of-arrival residuals from different pulsars. If the GW sources are of astrophysical origin, then they are located at discrete points on the sky. However, PTA data are often modeled, and subsequently analyzed, via a "standard Gaussian ensemble." That ensemble is obtained in the limit of an infinite density of vanishingly weak, Poisson-distributed sources. In this paper, we move away from that ensemble, to study the effects of two types of "source anisotropy." The first (a), which is often called "shot noise," arises because there are $N$ discrete GW sources at specific sky locations. The second (b) arises because the GW source positions are not a Poisson process, for example, because galaxy locations are clustered. Here, we quantify the impact of (a) and (b) on the mean and variance of the pulsar-averaged Hellings and Downs correlation. For conventional PTA sources, we show that the effects of shot noise (a) are much larger than the effects of clustering (b).

DOI: 10.1103/PhysRevD.110.123507

## I. INTRODUCTION

As pulsar timing arrays (PTAs) work towards $5\sigma$ detections of gravitational waves (GWs) [1–4], there is growing interest in how they might be used to study important questions in cosmology and astrophysics.

PTA data is often modeled and analyzed by assuming that the GW sources form a standard Gaussian ensemble [5–11]. This ensemble is obtained from randomly placed discrete sources in the limit as the spatial density of sources goes to infinity, with the average strength of each source taken to zero, in a way that leaves the mean-squared GW strain a constant [12,13]. The resulting source ensemble is often called "purely isotropic."

Here, we investigate two approaches to constructing more realistic source models. First, by assuming that the sources are discrete, meaning that there is a finite number of them, at specific (but unknown) sky locations, and second, by assuming that the sources have correlations in their spatial or angular locations. Our approach complements prior work on the effects of source discreteness, e.g., [14–18].

The first type of anisotropy occurs in a purely Poisson process, for which the probability that a source is located in some small volume $dV$ is proportional to $dV$ and independent of where any other sources might be located. Its effects are inversely proportional to the number of point sources and vanish in the limit of an infinite number of sources. In the cosmic microwave background and galaxy structure literature, this is often called "shot noise" [19,20].

[*]Contact author: bruce.allen@aei.mpg.de
[†]Contact author: deepali.agarwal@uclouvain.be
[‡]Contact author: joseph.romano@utrgv.edu
[§]Contact author: serena.valtolina@aei.mpg.de

The second type of anisotropy arises from correlations in the positions of the different sources. Such correlations would arise, for example, if the GW sources were located in galaxies, since galaxies tend to clump into clusters and filaments [21].

The purpose of this paper is to assess the impact of these two types of anisotropy on the mean and (co)variance of Hellings and Downs correlation. We estimate the order-of-magnitude of the effects described above. A brief outline of the paper follows.

Our approach exploits recent work on modeling nontrivial source angular distributions/correlations [22,23]. That work employs an ensemble of functions $\psi(\Omega)$ on the two-sphere. In Sec. II, we construct a rotationally invariant ensemble of functions whose support is at $N$ discrete points on the sphere. We then compute the first $\langle \psi_{lm} \rangle$ and second $\langle \psi_{lm} \psi_{l'm'} \rangle$ moments of $\psi$'s harmonic expansion coefficients, and the correlation function $\langle \psi(\Omega) \psi(\Omega') \rangle$ and covariance $C(\Omega, \Omega') \equiv \langle \psi(\Omega) \psi(\Omega') \rangle - \langle \psi(\Omega) \rangle \langle \psi(\Omega') \rangle$. In Sec. III we use these with the general framework developed in [22,23] to compute the cosmic variance for the discrete ensemble of $N$ sources. This provides a quantitative measure of how the discreteness of GW sources affects the Hellings and Downs correlation. In Sec. IV, we derive the same cosmic variance with a different but equivalent approach. Rather than employing a masking function, we instead construct an ensemble of discrete sources, whose amplitudes are drawn from a Gaussian ensemble. This demonstrates that our physical interpretation of the masking function is correct. In Sec. V, we extend this work by constructing an ensemble in which the number $N$ of discrete point sources is not fixed, but instead follows a Poisson distribution with mean $M$. Finally, in Sec. VI, we use these results to estimate the impact of GW source discreteness and correlations on current GW searches and on our ability to reconstruct the Hellings and Downs correlation. For the expected PTA sources, the effects of shot noise are small, but still large enough that they dominate the effects of galaxy structure correlations. (The same is true for the current generation of ground-based audio-band GW detectors, though that may change with the next generation.) This is followed by a short conclusion.

Note that the meaning of "shot noise" depends upon the context. The term "Schroteffekt" was originally coined by Schottky [24] to describe the discrete fluctuations in current due to individual electrons; the German word "Schrot" refers to small pellets or shot, as used in ammunition. It was later applied more broadly by Rice [25] [Eq. (1.5-1)] to describe a sum $\sum_j a_j f(t - t_j)$ of identical functions of time $f(t)$, scaled in amplitude and displaced in time. Typically, the $t_j$ are Poisson distributed, and the $a_j$ are Gaussian distributed. These can be used, for example, to model random fluctuations in pulsar pulse arrival times [26]. In this paper, "shot noise" refers to the statistical effects that arise because, as seen from Earth, gravitational wave sources are located at specific points on the sky. So, in our context, shot noise means a discrete spatial point process on the two-sphere, as is used in the literature on cosmological structure [21,27]. The term "shot noise" has also been used in this way to describe the temporal and spatial statistics of GW sources for ground-based audio-band detectors such as LIGO and Virgo [28,29].

## II. AN ENSEMBLE OF MASKING FUNCTIONS $\psi(\Omega)$ ON THE TWO-SPHERE

Recent work [22,23,30] has provided a formalism for computing the mean, variance, and covariance of the Hellings and Downs correlation for anisotropic source distributions. These can encompass both the effects of having a discrete set of sources, and the effects of having correlations in those source locations.

We begin by using those methods to model the effects of discreteness in the source locations. Later, in Sec. IV, we show that identical results can be obtained by using the "discrete point source" formalism in [12,13], but assuming that the source amplitudes have a Gaussian distribution.

To apply the methods of [22,23], we need to construct an ensemble of functions $\psi(\Omega)$. These are functions on the sphere, which encode the "discreteness" of the source locations.

### A. One masking function $\psi(\Omega)$ for $N$ points on the sphere

We start by constructing a single function $\psi(\Omega)$, to encode the positions of $N$ points, located at positions $\Omega_j$ on the unit two-sphere for $j = 1, ..., N$. This provides an exact and rigorous mathematical description of one instance or realization of a shot noise process. The (real) masking function is

$$\psi(\Omega) \equiv \frac{4\pi}{N} \sum_{j=1}^{N} \delta^2(\Omega, \Omega_j). \qquad (2.1)$$

Note that with this overall $4\pi/N$ normalization factor,

$$\int d\Omega \, \psi(\Omega) = 4\pi, \qquad (2.2)$$

where the integral over the two sphere has the usual solid-angle measure $d\Omega \equiv \sin\theta \, d\theta d\phi$.

The two-dimensional delta function on the sphere may be written (or alternatively defined) as

$$\delta^2(\Omega, \Omega') = \sum_{l=0}^{\infty} \sum_{m=-l}^{l} Y_{lm}(\Omega) Y_{lm}^*(\Omega'), \qquad (2.3)$$

where the $Y_{lm}$ are normal (scalar, spin-weight zero) spherical harmonics on the sphere. Going forward we will indicate sums of this form with $\sum_{lm}$.

The harmonic transform of $\psi$ can now be obtained by inserting (2.3) into (2.1). We obtain

$$\psi(\Omega) = \sum_{lm} \psi_{lm} Y_{lm}(\Omega), \qquad (2.4)$$

where the harmonic coefficients are

$$\psi_{lm} = \frac{4\pi}{N} \sum_{j=1}^{N} Y_{lm}^*(\Omega_j). \qquad (2.5)$$

We define the rotationally invariant coefficients

$$\begin{aligned}\psi_l &\equiv \frac{1}{2l+1} \sum_{m=-l}^{l} |\psi_{lm}|^2 \\ &= \frac{(4\pi)^2}{N^2(2l+1)} \sum_{j=1}^{N} \sum_{k=1}^{N} \sum_{m=-l}^{l} Y_{lm}(\Omega_j) Y_{lm}^*(\Omega_k) \\ &= \frac{4\pi}{N^2} \sum_{j=1}^{N} \sum_{k=1}^{N} P_l(\Omega_j \cdot \Omega_k) \\ &= \frac{4\pi}{N}\left[1 + \frac{1}{N} \sum_{j \neq k} P_l(\Omega_j \cdot \Omega_k)\right],\end{aligned} \qquad (2.6)$$

where we used the addition theorem for spherical harmonics to go from the second to the third line. On the final line, we have broken the sum into diagonal terms for which $j = k$ and off-diagonal terms for which $j \neq k$, then used $P_l(1) = 1$. This and the previous equations hold for any set of $N$ points on the sphere.

### B. A rotationally invariant ensemble of $\psi(\Omega)$

We now consider a rotationally invariant ensemble of functions $\psi(\Omega)$. This provides an exact and rigorous mathematical description of a "statistically rotationally invariant" shot noise process on the sphere. The ensemble of functions $\psi(\Omega)$ is constructed by taking a large number of realizations, each of which has $N$ points distributed at random, uniformly and independently, on the sphere. Angle brackets now refer to averages over that ensemble of different functions or equivalently, different choices of point locations on the sphere. Every realization in the ensemble has the same number $N$ of points.

The first moment of the harmonic coefficients (2.5) is

$$\langle \psi_{lm} \rangle = \frac{4\pi}{N} \sum_{j=1}^{N} \langle Y_{lm}^*(\Omega_j) \rangle. \qquad (2.7)$$

Since the expected value is an average over all realizations of $N$ points randomly selected on the sphere, the right-hand side is a Monte Carlo approximation of the average value of the spherical harmonic function over the sphere. For a large number of realizations we thus have, independent of $N$, that

$$\frac{1}{N} \sum_{j=1}^{N} \langle Y_{lm}^*(\Omega_j) \rangle = \int \frac{d\Omega}{4\pi} Y_{lm}^*(\Omega) = \frac{1}{\sqrt{4\pi}} \delta_{l0} \delta_{m0}; \qquad (2.8)$$

the only case which does not integrate to zero is if $l = m = 0$, for which the spherical harmonic is a constant. Thus, combining (2.7) and (2.8) gives

$$\langle \psi_{lm} \rangle = \sqrt{4\pi} \delta_{l0} \delta_{m0}. \qquad (2.9)$$

Note that (2.9) implies

$$\langle \psi(\Omega) \rangle = 1, \qquad (2.10)$$

which is also consistent with our normalization condition on $\psi$ given in (2.2). This is also the normalization condition given in [22] [Eq. (9.2)].

The second moment of the coefficients is not needed (it is enough to work with $\psi_l$) but since it is easy to compute, we do that also. It is helpful to first compute the ensemble average of $Y_{lm}(\Omega_j) Y_{l'm'}^*(\Omega_k)$. Since the $\Omega_j$ are chosen independently, if $j \neq k$ then this simply factors to give

$$\begin{aligned}\text{for } j \neq k \colon \langle Y_{lm}(\Omega_j) Y_{l'm'}^*(\Omega_k) \rangle &= \langle Y_{lm}(\Omega_j) \rangle \langle Y_{l'm'}^*(\Omega_k) \rangle \\ &= \frac{1}{4\pi} \delta_{ll'} \delta_{mm'} \delta_{l0} \delta_{m0}.\end{aligned} \qquad (2.11)$$

Here, we have used the ensemble average $\langle Y_{lm}(\Omega_j) \rangle = \delta_{l0} \delta_{m0}/\sqrt{4\pi}$, which follows immediately from (2.8), since every term that appears in the sum must have the same value. For $j = k$ the arguments of the spherical harmonics are equal, and we obtain

$$\begin{aligned}\text{for } j = k \colon \langle Y_{lm}(\Omega_j) Y_{l'm'}^*(\Omega_k) \rangle &= \int \frac{d\Omega}{4\pi} Y_{lm}(\Omega) Y_{l'm'}^*(\Omega) \\ &= \frac{1}{4\pi} \delta_{ll'} \delta_{mm'},\end{aligned} \qquad (2.12)$$

which follows from the orthonormality of the spherical harmonic functions on the sphere. If desired, the formulas for the cases $j \neq k$ and $j = k$ can be combined into a single form which is valid for all $j$ and $k$:

$$\langle Y_{lm}(\Omega_j) Y_{l'm'}^*(\Omega_k) \rangle = \frac{1}{4\pi} \delta_{ll'} \delta_{mm'} \left[ \delta_{jk} + (1 - \delta_{jk}) \delta_{l0} \delta_{m0} \right]. \qquad (2.13)$$

With this, it is trivial to compute the second moment of the coefficients $\psi_{lm}$.

From (2.5), it follows immediately that the second moments are

$$\langle \psi_{lm}\psi^*_{l'm'}\rangle = \left(\frac{4\pi}{N}\right)^2 \sum_{j=1}^{N}\sum_{k=1}^{N} \langle Y^*_{lm}(\Omega_j) Y_{l'm'}(\Omega_k)\rangle$$

$$= \left(\frac{4\pi}{N}\right)^2 \left[\frac{N}{4\pi}\delta_{ll'}\delta_{mm'} + \frac{N(N-1)}{4\pi}\delta_{ll'}\delta_{mm'}\delta_{l0}\delta_{m0}\right]$$

$$= \frac{4\pi}{N}\left[1 + (N-1)\delta_{l0}\delta_{m0}\right]\delta_{ll'}\delta_{mm'}, \quad (2.14)$$

where we have broken the sum into $N$ diagonal terms for which $j = k$ and we used (2.12), and $N(N-1)$ off diagonal terms for which we used (2.11). The correlation function is then

$$\langle \psi(\Omega)\psi(\Omega')\rangle = \sum_{lm}\sum_{l'm'}\langle \psi_{lm}\psi^*_{l'm'}\rangle Y_{lm}(\Omega)Y^*_{l'm'}(\Omega')$$

$$= \frac{4\pi}{N}\sum_{lm}\left[1 + (N-1)\delta_{l0}\delta_{m0}\right]Y_{lm}(\Omega)Y^*_{lm}(\Omega')$$

$$= \frac{4\pi}{N}\delta^2(\Omega,\Omega') + \frac{N-1}{N}. \quad (2.15)$$

The covariance [22] [Eq. (9.3)] is

$$C(\Omega,\Omega') \equiv \langle \psi(\Omega)\psi(\Omega')\rangle - \langle \psi(\Omega)\rangle\langle \psi(\Omega')\rangle$$

$$= \frac{4\pi}{N}\delta^2(\Omega,\Omega') - \frac{1}{N}, \quad (2.16)$$

where we have used (2.10) to obtain the second equality. Note that both sides can be written as a function of $\Omega \cdot \Omega'$: returning to (2.15) and using the addition theorem, we can write the correlation function in the form

$$\langle \psi(\Omega)\psi(\Omega')\rangle = 1 + \sum_{l=1}^{\infty}\frac{2l+1}{N}P_l(\Omega \cdot \Omega'). \quad (2.17)$$

The covariance then takes the form

$$C(\Omega,\Omega') = \sum_{l=1}^{\infty}\frac{2l+1}{N}P_l(\Omega \cdot \Omega'). \quad (2.18)$$

Using the notation of [22] [Eq. (9.4)] this corresponds to expansion coefficients

$$C_l = \begin{cases} 0 & \text{for } l = 0 \\ 4\pi/N & \text{for } l > 0. \end{cases} \quad (2.19)$$

Note that it is only in the limit $N \to \infty$ that this ensemble of Poisson-distributed point sources has all $C_l$ vanish, corresponding to the standard Gaussian source ensemble.

We can obtain an identical result by computing the ensemble average of $\psi_l$ given in (2.6). This follows immediately from (2.6) because

$$\frac{1}{2}\int_{-1}^{1}dz\,P_l(z) = \delta_{l0}. \quad (2.20)$$

This implies that the ensemble average of $P_l(\Omega_j \cdot \Omega_k)$ for $j \neq k$ is unity for $l = 0$ and vanishes for $l > 0$. Hence,

$$\langle \psi_l\rangle = \begin{cases} 4\pi & \text{for } l = 0 \\ 4\pi/N & \text{for } l > 0. \end{cases} \quad (2.21)$$

This is consistent with the $C_l$ given in (2.19), since $C_l = \langle \psi_l\rangle$ for $l > 0$ and $C_0 = \langle \psi_0\rangle - 4\pi$.

## III. EFFECTS OF ANISOTROPY ARISING FROM THE DISCRETENESS OF THE GW SOURCE LOCATIONS

We now use the results of the previous section to compute the effects of the discreteness of the source locations on the mean, cosmic variance, and cosmic covariance of the Hellings and Downs correlation. For this, we use the formalism of [22,23]. To follow the details of the calculations below, the reader must consult Sec. IX of [22].

The quantity $\Gamma(\gamma)$ denotes the pulsar-averaged Hellings and Downs correlation at angle $\gamma$, and $h^2$ and $\hbar^2$ are measures of the mean-squared GW strain at Earth. These are defined in terms of the GW power spectrum $H(f)$ by Eqs. (C19) and (C26) of [12]. Additional discussion can be found in Appendices A and B of [10].

The ensemble average of the pulsar-averaged correlation is

$$\langle \Gamma(\gamma)\rangle_\psi = h^2\mu_u(\gamma). \quad (3.1)$$

[In this section, we use the $\langle \rangle_\psi$ notation from [22,23] to indicate the average over a set of Gaussian subensembles labeled by masking functions. This is indicated *without* a subscript in Sec. II, and differs from the standard Gaussian ensemble average. The meaning of $\langle \rangle$ without a subscript is context dependent.] This is proportional to the Hellings and Downs curve $\mu_u(\gamma)$ and is the same as for the standard Gaussian ensemble. The deviation of any representative of the ensemble away from this mean is

$$\Delta\Gamma(\gamma) \equiv \Gamma(\gamma) - \langle \Gamma(\gamma)\rangle_\psi. \quad (3.2)$$

It follows by definition that the ensemble average of $\Delta\Gamma$ vanishes: $\langle \Delta\Gamma(\gamma)\rangle_\psi = 0$. The cosmic variance is

$$\sigma^2_{\text{cos}}(\gamma) \equiv \langle (\Delta\Gamma(\gamma))^2\rangle_\psi = \langle (\Gamma(\gamma))^2\rangle_\psi - \langle \Gamma(\gamma)\rangle^2_\psi, \quad (3.3)$$

whereas the cosmic covariance is

$$\sigma^2_{\text{cos}}(\gamma,\gamma') \equiv \langle \Delta\Gamma(\gamma)\Delta\Gamma(\gamma')\rangle_\psi$$

$$= \langle \Gamma(\gamma)\Gamma(\gamma')\rangle_\psi - \langle \Gamma(\gamma)\rangle_\psi\langle \Gamma(\gamma')\rangle_\psi. \quad (3.4)$$

It follows by definition that the cosmic variance is the diagonal part of the cosmic covariance: $\sigma^2_{\text{cos}}(\gamma) = \sigma^2_{\text{cos}}(\gamma,\gamma)$.

While the ensemble average (3.1) of $\Gamma(\gamma)$ is the same as for the standard Gaussian ensemble, the cosmic variance and covariance are different. Taking the ensemble average of [22] [Eq. (9.8)], and denoting the angle between $\Omega$ and $\Omega'$ by $\beta$, we obtain

$$\langle\Gamma(\gamma)\Gamma(\gamma')\rangle_\psi = \int \frac{d\Omega}{4\pi} \int \frac{d\Omega'}{4\pi} \langle\psi(\Omega)\psi(\Omega')\rangle_\psi \Big[h^4 \mu_u(\gamma)\mu_u(\gamma')$$
$$+ \hbar^4\big(\mu(\gamma,\beta)\mu(\gamma',\beta) + \mu(\bar\gamma,\bar\beta)\mu(\bar{\gamma'},\bar\beta)\big)\Big]$$
$$= \left(h^4 + \frac{1}{N}\hbar^4\right)\mu_u(\gamma)\mu_u(\gamma')$$
$$+ 2\hbar^4\left(\frac{N-1}{N}\right)\tilde\mu^2(\gamma,\gamma'), \quad (3.5)$$

where to obtain the final equality we have substituted (2.15) and used this to do the integrals. Here $\bar\gamma = \pi - \gamma$, $\bar{\gamma'} = \pi - \gamma'$, and $\bar\beta = \pi - \beta$.

The Hellings and Downs curve $\mu_u(\gamma)$ and the two-point function $\mu(\gamma,\beta)$ are defined and discussed in [12,22]. To evaluate the integrals we have used two properties of the two-point function. First, that for coincident points it reduces to the Hellings and Downs curve: $\mu(\gamma,0) = \mu_u(\gamma)$, and second, that for antipodal GW source points, it vanishes: $\mu(\gamma,\pi) = 0$. To evaluate the second integral we have used $\tilde\mu^2(\gamma,\gamma')$ as defined by [22] [Eq. (7.17) or Eq. (8.4)].

Returning to the cosmic covariance, the second term in (3.5) is proportional to the cosmic covariance for the standard Gaussian ensemble [22] [Eq. (7.16)], which is

$$\sigma^2_{\text{cos,std}}(\gamma,\gamma') = 2\hbar^4 \tilde\mu^2(\gamma,\gamma'). \quad (3.6)$$

Thus, subtracting the mean (3.1) from (3.5) gives the cosmic covariance

$$\langle\Delta\Gamma(\gamma)\Delta\Gamma(\gamma')\rangle_\psi \equiv \langle\Gamma(\gamma)\Gamma(\gamma')\rangle_\psi - \langle\Gamma(\gamma)\rangle_\psi\langle\Gamma(\gamma')\rangle_\psi$$
$$= \frac{1}{N}\hbar^4 \mu_u(\gamma)\mu_u(\gamma')$$
$$+ \left(1 - \frac{1}{N}\right)\sigma^2_{\text{cos,std}}(\gamma,\gamma'). \quad (3.7)$$

If we let $\gamma = \gamma'$, then we get the cosmic variance

$$\sigma^2_{\text{cos}}(\gamma) = \frac{1}{N}\hbar^4 \mu_u^2(\gamma) + \left(1 - \frac{1}{N}\right)\sigma^2_{\text{cos,std}}(\gamma). \quad (3.8)$$

Thus, the cosmic variance and covariance for a discrete set of $N$ point sources, obtained by "masking" the standard Gaussian ensemble, differ from those of the standard Gaussian ensemble. The difference terms are proportional to $1/N$. Only in the $N \to \infty$ limit does the masked result agree with that of the standard Gaussian ensemble.

This result can also be obtained by substituting the $C_l$ given in (2.21) into [22] [Eq. (9.13)] or [23] [Eq. (4.23)].

## IV. COMPARISON WITH AN ENSEMBLE OF DISCRETE SOURCES

Recent work [12,13] calculates the mean and variance of the Hellings and Downs correlation for a set of $N$ discrete point sources. The ensembles used in the two citations differ in their assumptions about the polarization properties. However, both assume that the $j$th GW source has exactly the same average-squared amplitude in every realization in the ensemble. Here, we undertake a similar construction, but assume that the amplitudes *differ from one realization to the next* with a Gaussian distribution. We show that this ensemble exactly reproduces the "masked" results of Sec. III.

We begin with $N$ arbitrary-waveform GW sources in a single realization of the universe. These are located at specific directions $-\hat\Omega_j$ on the sky, where $j = 1, \ldots N$ labels the sources. The $j$th source has an arbitrary time-domain strain in the $A = +, \times$ polarizations given by

$$h_j^A(t) = \int_{-\infty}^\infty df\, \tilde h_j^A(f)\, e^{i2\pi f t}. \quad (4.1)$$

The Fourier amplitudes $\tilde h_j^A(f)$ satisfy $\tilde h_j^A(-f) = \tilde h_j^{A*}(f) \equiv (\tilde h_j^A(f))^*$, which ensures that the strains in (4.1) are real. Provided that they satisfy this constraint, the Fourier amplitudes $\tilde h_j^A(f)$ may be arbitrary functions of frequency.

The (real) metric perturbation that results is a sum over the $N$ sources, given by

$$h_{ab}(t,\mathbf{x}) \equiv \sum_{j=1}^N \sum_{A=+,\times} h_j^A(t - \hat\Omega_j \cdot \mathbf{x})\, e_{ab}^A(\Omega_j)$$
$$= \sum_j \sum_A \int df\, \tilde h_j^A(f)\, e^{i2\pi f(t-\hat\Omega_j\cdot\mathbf{x})}\, e_{ab}^A(\Omega_j). \quad (4.2)$$

We have taken $\mathbf{x} = \mathbf{0}$ at the Earth/Solar System Barycenter, and $e_{ab}^A(\Omega_j)$ denote polarization tensors defined in Eq. (D6) of [12]. Starting from the second equality above, unless indicated otherwise, sums over sources (labeled by $j, k, \ldots$) are from 1 to $N$ and sums over polarizations (labeled by $A, A', \ldots$) are over $+, \times$. Similarly, integrals over frequency $f, f', \ldots$ are over the range $(-\infty, \infty)$ unless otherwise indicated. Note that $h_{ab}(t,\mathbf{x})$ satisfies the wave equation and is transverse, traceless, and synchronous.

The effect of the GW on pulsar redshift is described in [11,12]. The redshift of pulsar $p$ as observed at Earth at time $t$ is

$$Z_p(t) = \sum_j \sum_A \int df\, \tilde h_j^A(f)\, e^{i2\pi f t} F_p^A(\Omega_j)$$
$$\times \left[1 - e^{-i2\pi f L_p(1+\hat\Omega\cdot\hat p)}\right]. \quad (4.3)$$

Here, $L_p$ is the light travel time from Earth to pulsar, $\hat p$ is a unit vector from Earth to pulsar, and $F_p^A(\Omega_j)$ is the antenna pattern function for pulsar $p$, defined in Eq. (2.1) of [12]. The so-called "Earth term" arises from the "1" in square brackets, and the "pulsar term" arises from the pure-phase exponential [11] [Eq. (23)].

The correlation between two pulsars $p$ and $q$ is the time-averaged product of their redshifts, which we denote $\overline{Z_p Z_q}$ (but see note added in proof). Averaging this product over the observation time interval $t \in [-T/2, T/2]$ introduces a sinc function as given below. The pulsar-averaged correlation $\Gamma(\gamma)$ is obtained from this by additionally averaging over all pairs of pulsars at angular separation $\gamma$:

$$\Gamma(\gamma) \equiv \langle \overline{Z_p Z_q} \rangle_{pq \in \gamma}. \quad (4.4)$$

The notation $\langle Q \rangle_{pq \in \gamma}$ indicates the average of $Q$ over all pulsar pairs $p$, $q$, uniformly distributed over the sky, with angular separation $\gamma$.

This pulsar-averaged correlation may be computed from (4.3) and (4.4) as shown in [12]. The pulsar-average of the antenna pattern functions is called the two-point function. It is denoted

$$\mu_{AA'}(\gamma, \Omega_j, \Omega_k) \equiv \langle F_p^A(\Omega_j) F_q^{A'}(\Omega_k) \rangle_{pq \in \gamma} \quad (4.5)$$

and is computed explicitly in [22]. Properly taking into account the complex phase of $\mu$ (described in [22], where it is denoted $\chi$) and using the fact that $Z_q$ is real, so that it can be replaced by its complex conjugate, we obtain

$$\Gamma(\gamma) = \sum_{j,k} \sum_{A,A'} \int df \int df' \, \tilde{h}_j^A(f) \tilde{h}_k^{A'*}(f') \mu_{AA'}(\gamma, \Omega_j, \Omega_k)$$
$$\times \operatorname{sinc}(\pi(f - f')T). \quad (4.6)$$

Only the Earth terms survive the pulsar average (see [12] for more details). For the sinc function we use the "mathematics" definition $\operatorname{sinc}(x) = (\sin x)/x$, which has its first zero at $x = \pi$. Note that all of the $\Omega_j$ dependence of $\Gamma$ is via the two-point functions $\mu_{AA'}$, whereas all of the dependence upon the amplitudes $\tilde{h}_j^A(f)$ is outside of the two-point functions.

Up to here, all of our equations describe a single realization of the universe, in which the $N$ GW sources have specific (but arbitrary) directions and waveforms. To study the statistical properties of the GW background, we now construct an ensemble of different universes, consisting of many realizations, with different source waveforms and directions. Because we do not know the values assumed by those quantities in our own universe, we can employ the ensemble to make statements about the likelihood of certain outcomes or measurements.

Each realization in the ensemble contains $N$ GW sources. Each realization is defined by $2N$ complex Fourier amplitudes $\tilde{h}_j^{+,\times}(f)$ and by source directions $\Omega_j$, for $j = 1, \ldots N$. In each of these realizations, (4.1) and (4.2) define the strain of the GW emission from the $j$th source, and (4.6) is the pulsar-averaged pulsar correlation at angle $\gamma$.

In the ensemble, we will pick the source directions $\Omega_j$ uniformly and independently on the two-sphere. The $\tilde{h}_j^{+,\times}(f)$ are chosen from a Gaussian ensemble which is independent for each source, has vanishing mean, and has second moment (power spectrum) $\mathcal{H}(f)$. [Note that $\mathcal{H}(f)$ is analogous to the power spectrum $H(f)$, but for a single source; see discussion before (4.10).] This means that the ensemble average of any functional $Q$ of the GW strain can be computed in two steps. First, one averages $Q$ over the Gaussian ensemble of Fourier amplitudes, and second, one averages over the source positions. Symbolically, this is

$$\langle Q \rangle \equiv \prod_{j=1}^N \left( \int \frac{d\Omega_j}{4\pi} \right) \prod_{k=1}^N \left[ \int d[\tilde{h}_k^+] e^{-(\tilde{h}_k^+, \tilde{h}_k^+)/2} \int d[\tilde{h}_k^\times] e^{-(\tilde{h}_k^\times, \tilde{h}_k^\times)/2} \right] Q(\tilde{h}_1^+(f), \ldots, \tilde{h}_N^\times(f), \Omega_1 \ldots \Omega_N), \quad (4.7)$$

where the measure on the functions is normalized so that $\langle 1 \rangle = 1$. The positive-definite inner product, which appears in (4.7), is

$$(A, B) = \int df \, \frac{A(f) B^*(f)}{\mathcal{H}(f)}, \quad (4.8)$$

where $\mathcal{H}(f)$ is a non-negative function whose interpretation is given below.

We use the angle brackets to denote the full ensemble average, and add a subscript to indicate an average over only some of the random variables. The subscript $\Omega$ refers to an average over the sky positions, and the subscript $\mathcal{A}$ indicates an average over the Gaussian ensemble of amplitudes. Thus, $\langle Q \rangle \equiv \langle Q \rangle_{\mathcal{A}, \Omega}$.

In this paper, the functionals $Q$ of interest are quadratic or quartic functions of the Fourier amplitudes. So in practice, to evaluate the averages over these amplitudes, we use the first and second moments

$$\langle \tilde{h}_j^A(f) \rangle_{\mathcal{A}} = 0,$$
$$\langle \tilde{h}_j^A(f) \tilde{h}_k^{A'*}(f') \rangle_{\mathcal{A}} = 4\pi \mathcal{H}(f) \delta(f - f') \delta_{jk} \delta_{AA'}. \quad (4.9)$$

Since these variables are Gaussian, higher moments with averages over the amplitudes $\mathcal{A}$ can be computed via Isserlis's theorem [31]. Note, however, that the full ensemble is not Gaussian: Isserlis's theorem can only be applied to averages over the Gaussian amplitudes.

The function $\mathcal{H}(f)$ is the spectral distribution of GW power for one source; the overall factor in (4.9) is selected for consistency with previous results and standard literature. In a universe containing $N$ sources, the standard spectral function $H(f)$ (as used for example in Sec. III) is $H(f) = N\mathcal{H}(f)$.

Computing the first moment $\langle\Gamma(\gamma)\rangle$ is straightforward. Taking the ensemble average of (4.6), we obtain

$$\begin{aligned}\langle\Gamma(\gamma)\rangle &= \sum_{j,k}\sum_{A,A'}\int df\int df'\,\text{sinc}(\pi(f-f')T)\langle\tilde{h}_j^A(f)\tilde{h}_k^{A'*}(f')\rangle_\mathcal{A}\langle\mu_{AA'}(\gamma,\Omega_j,\Omega_k)\rangle_\Omega \\ &= \sum_j\sum_A\int df\,4\pi\,\mathcal{H}(f)\langle\mu_{AA}(\gamma,\Omega_j,\Omega_j)\rangle_\Omega \\ &= h^2\mu_\text{u}(\gamma).\end{aligned} \quad (4.10)$$

The first equality follows from the ensemble average of (4.6), the second follows from (4.7), (4.9), and $\text{sinc}(0) = 1$, the third follows by definition of $h^2$ (immediately below), and because $\sum_A \mu_{AA}(\gamma,\Omega,\Omega) = \mu_\text{u}(\gamma)$ is the Hellings and Downs curve (see Appendix C of [22] for more details). We have defined

$$h^2 \equiv 4\pi N \int df\,\mathcal{H}(f), \quad (4.11)$$

which is a measure of the mean-squared GW strain at Earth (see Eq. (C19) of [12]). Because the different sources are statistically uncorrelated, this is $N$ times larger than the mean-squared strain of a single source.

To compute the cosmic covariance, we first evaluate the ensemble average $\langle\Gamma(\gamma)\Gamma(\gamma')\rangle$. Again, all of the amplitude dependence (indicated by the subscript $\mathcal{A}$) is via $\tilde{h}_j^A(f)$; all of the dependence on source sky locations (indicated by the subscript $\Omega$) is via the two-point functions. Thus, we can write

$$\begin{aligned}\langle\Gamma(\gamma)\Gamma(\gamma')\rangle = \sum_{j,k,\ell,m}\sum_{A,A',A'',A'''}\int df\int df'\int df''\int df'''\,\text{sinc}(\pi(f-f')T)\text{sinc}(\pi(f''-f''')T) \\ \times \langle\tilde{h}_j^A(f)\tilde{h}_k^{A'*}(f')\tilde{h}_\ell^{A''}(f'')\tilde{h}_m^{A'''*}(f''')\rangle_\mathcal{A}\langle\mu_{AA'}(\gamma,\Omega_j,\Omega_k)\mu_{A''A'''}(\gamma',\Omega_\ell,\Omega_m)\rangle_\Omega.\end{aligned} \quad (4.12)$$

Since the $\tilde{h}_j^A(f)$ are Gaussian random variables, the fourth moment can be directly obtained from Isserlis's theorem [31]:

$$\begin{aligned}\langle\tilde{h}_j^A(f)\tilde{h}_k^{A'*}(f')\tilde{h}_\ell^{A''}(f'')\tilde{h}_m^{A'''*}(f''')\rangle_\mathcal{A} &= (4\pi)^2\mathcal{H}(f)\mathcal{H}(f'')\delta(f-f')\delta(f''-f''')\delta_{jk}\delta_{\ell m}\delta_{AA'}\delta_{A''A'''} \\ &+ (4\pi)^2\mathcal{H}(f)\mathcal{H}(f')\delta(f+f'')\delta(f'+f''')\delta_{j\ell}\delta_{km}\delta_{AA''}\delta_{A'A'''} \\ &+ (4\pi)^2\mathcal{H}(f)\mathcal{H}(f')\delta(f-f''')\delta(f'-f'')\delta_{jm}\delta_{k\ell}\delta_{AA'''}\delta_{A'A''}.\end{aligned} \quad (4.13)$$

Substituting (4.13) in (4.12), we obtain

$$\begin{aligned}\langle\Gamma(\gamma)\Gamma(\gamma')\rangle &= (4\pi)^2\int df\int df'\,\mathcal{H}(f)\mathcal{H}(f')\sum_{j,k}\sum_{A,A'}\langle\mu_{AA}(\gamma,\Omega_j,\Omega_j)\mu_{A'A'}(\gamma',\Omega_k,\Omega_k)\rangle_\Omega \\ &+ (4\pi)^2\int df\int df'\,\mathcal{H}(f)\mathcal{H}(f')\,\text{sinc}^2(\pi(f-f')T) \\ &\times \sum_{j,k}\sum_{A,A'}\langle\mu_{AA'}(\gamma,\Omega_j,\Omega_k)\mu_{AA'}(\gamma',\Omega_j,\Omega_k) + \mu_{AA'}(\gamma,\Omega_j,\Omega_k)\mu_{A'A}(\gamma',\Omega_k,\Omega_j)\rangle_\Omega,\end{aligned} \quad (4.14)$$

where we have relabeled some integration and summation variables. To complete the ensemble average, we need to evaluate the average over the source directions $\Omega$. These averages do not depend upon the values of $j$ and $k$, but only on whether or not $j$ and $k$ are the same or different.

The first $\Omega$ average in (4.14) is trivial. Since $\sum_A \mu_{AA}(\gamma, \Omega_j, \Omega_j) = \mu_u(\gamma)$, which is independent of the source direction $\Omega_j$,

$$\sum_{A,A'} \langle \mu_{AA}(\gamma, \Omega_j, \Omega_j) \mu_{A'A'}(\gamma', \Omega_k, \Omega_k) \rangle_\Omega = \mu_u(\gamma) \mu_u(\gamma'). \tag{4.15}$$

Note that for $\gamma = \gamma'$, the first term in (4.14) is the square of the first moment (4.10).

The ensemble average in the final line of (4.14) can be simplified in two steps. First, it follows from [22] [Eq. (C4)] that $\mu_{AA'}(\gamma, \Omega, \Omega') = \mu_{A'A}(\gamma, \Omega', \Omega)$, so the two terms are equal. Second, from [22] [Eq. (C4)], one can show that

$$\sum_{AA'} \mu_{AA'}(\gamma, \Omega_j, \Omega_k) \mu_{AA'}(\gamma', \Omega_j, \Omega_k)$$
$$= \frac{1}{2} [\mu(\gamma, \beta_{jk})\mu(\gamma', \beta_{jk}) + \mu(\bar\gamma, \bar\beta_{jk})\mu(\bar{\gamma'}, \bar\beta_{jk})]. \tag{4.16}$$

Here, $\cos\beta_{jk} \equiv \Omega_j \cdot \Omega_k$, $\bar\beta_{jk} \equiv \pi - \beta_{jk}$, $\bar\gamma = \pi - \gamma$, and $\bar{\gamma'} = \pi - \gamma'$.

To evaluate the average over $\Omega$, we must consider $j = k$ and $j \neq k$ separately. If $j = k$, then the right-hand side of (4.16) reduces to $\mu_u(\gamma)\mu_u(\gamma')/2$, which is unchanged by the $\Omega$ average. If $j \neq k$, then the $\Omega$ average of (4.16) is evaluated in [22] and gives the function $\tilde\mu^2(\gamma, \gamma')$, defined by

$$\tilde\mu^2(\gamma, \gamma') \equiv \int \frac{d\Omega_j}{4\pi} \int \frac{d\Omega_k}{4\pi} \mu(\gamma, \beta_{jk})\mu(\gamma', \beta_{jk})$$
$$= \frac{1}{2} \int_{-1}^{1} d(\cos\beta) \mu(\gamma, \beta) \mu(\gamma', \beta)$$
$$= \sum_{l=2}^{\infty} q_l P_l(\cos\gamma) P_l(\cos\gamma'). \tag{4.17}$$

Here, $P_l(\cos\gamma)$ are Legendre polynomials and the coefficients are [22] [(8.4), (2.11)]

$$q_l \equiv \frac{2l+1}{(l+2)^2(l+1)^2 l^2 (l-1)^2}. \tag{4.18}$$

[Note: in the notation of [22], $q_l = a_l^2/(2l+1)$.] Since $P_l(\cos\bar\gamma) = (-1)^l P_l(\cos\gamma)$, from the last equality in (4.17) one can see that $\tilde\mu^2(\gamma, \gamma') = \tilde\mu^2(\bar\gamma, \bar{\gamma'})$.

Combining these results gives a simple form for the ensemble average:

$$\langle \Gamma(\gamma) \Gamma(\gamma') \rangle = h^4 \mu_u(\gamma) \mu_u(\gamma') + \frac{1}{N} \hbar^4 \mu_u(\gamma) \mu_u(\gamma')$$
$$+ 2\left(1 - \frac{1}{N}\right) \hbar^4 \tilde\mu^2(\gamma, \gamma'), \tag{4.19}$$

where we used (4.15), (4.16), and (4.17) and have defined a measure of squared strain $\hbar^2$ by

$$\hbar^4 \equiv (4\pi)^2 N^2 \int df \int df' \mathcal{H}(f) \mathcal{H}(f') \text{sinc}^2(\pi(f-f')T). \tag{4.20}$$

Note that $\hbar^4$ is Eq. (C26) of [12] with $H(f) \equiv N\mathcal{H}(f)$.

Now we have all the elements to compute the cosmic covariance of the ensemble defined by (4.7). Starting from the definition (3.4) and using (4.10) and (4.19), the cosmic covariance is

$$\sigma_{\text{cos}}^2(\gamma, \gamma') = \langle \Gamma(\gamma) \Gamma(\gamma') \rangle - \langle \Gamma(\gamma) \rangle \langle \Gamma(\gamma') \rangle$$
$$= \frac{1}{N} \hbar^4 \mu_u(\gamma) \mu_u(\gamma') + 2\left(1 - \frac{1}{N}\right) \hbar^4 \tilde\mu^2(\gamma, \gamma')$$
$$= \frac{1}{N} \hbar^4 \mu_u(\gamma) \mu_u(\gamma') + \left(1 - \frac{1}{N}\right) \sigma_{\text{cos,std}}^2(\gamma, \gamma'), \tag{4.21}$$

where for the final equality we have used (3.6). This agrees exactly with the cosmic covariance (3.7) obtained in Sec. III for the masked ensemble. The cosmic variances are obtained as shown following (3.4), and thus are also in agreement: setting $\gamma = \gamma'$ in (4.21) reproduces (3.8).

In preparation for the following section, it is useful to define *single source* versions of $h^2$ and $\hbar^4$, which are denoted with a subscript "s". These are obtained by setting the number of GW sources to $N = 1$ in (4.11) and (4.20):

$$h_s^2 \equiv 4\pi \int df \mathcal{H}(f) = h^2/N,$$
$$\hbar_s^4 \equiv (4\pi)^2 \int df \int df' \mathcal{H}(f) \mathcal{H}(f') \text{sinc}^2(\pi(f-f')T)$$
$$= \hbar^4/N^2. \tag{4.22}$$

Since $h^2 = N h_s^2$ and $\hbar^4 = N^2 \hbar_s^4$, the expression for the cosmic covariance, given by the second equality in (4.21), becomes

$$\sigma_{\text{cos}}^2(\gamma, \gamma') = N \hbar_s^4 [\mu_u(\gamma) \mu_u(\gamma') + 2(N-1) \tilde\mu^2(\gamma, \gamma')]. \tag{4.23}$$

It immediately follows from this expression that

$$\sigma_{\text{cos}}^2(\gamma)|_{N=0} = 0, \quad \sigma_{\text{cos}}^2(\gamma)|_{N=1} = \hbar_s^4 \mu_u^2(\gamma),$$
$$\sigma_{\text{cos}}^2(\gamma)|_{N=2} = 2\hbar_s^4 [\mu_u^2(\gamma) + 2\tilde\mu^2(\gamma)]. \tag{4.24}$$

For no GW sources, there is no cosmic variance, because the pulsar-averaged correlation vanishes in every realization. For a universe containing a single source, in every realization the pulsar-averaged correlation has *exactly* the shape of the Hellings and Downs curve, since there is no source interference. However, the amplitude of the correlation will vary from one realization of the universe to another, because the amplitude of the source is drawn from a Gaussian distribution. Thus, one obtains a cosmic variance proportional to the square of the Hellings and Down curve, $\mu_u^2(\gamma)$. Once there are two or more GW sources, interference between the sources gives rise to the additional $\tilde{\mu}^2(\gamma)$ term, whereas the first term continues to describe the realization-to-realization dependent variations in the mean-squared strain at Earth. Figure 1 compares the contributions of $\mu_u^2(\gamma)$ and $\tilde{\mu}^2(\gamma)$ to the cosmic variance.

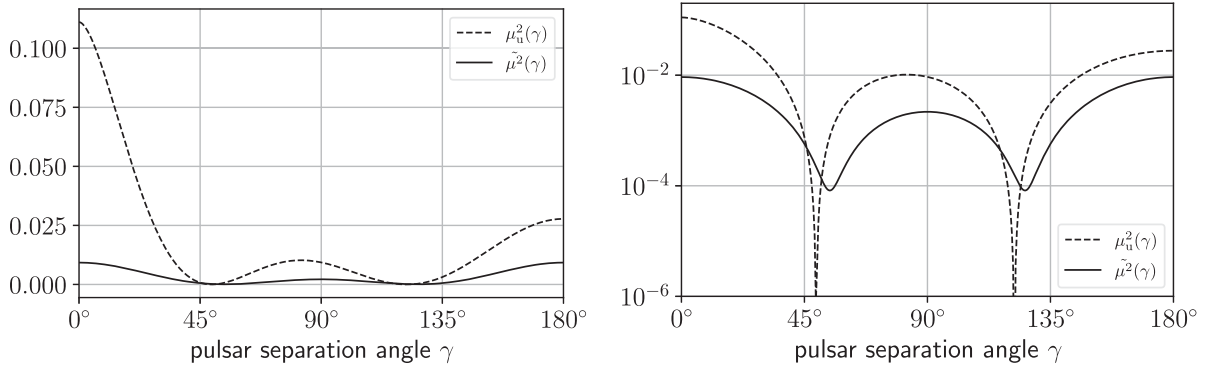

FIG. 1. Comparison of $\mu_u^2(\gamma)$ and $\tilde{\mu}^2(\gamma)$. Left panel: linear scale. Right panel: logarithmic scale, to better illustrate the regions around the two zeroes of the Hellings and Downs curve.

## V. POISSON DISTRIBUTION OF SOURCES

We have constructed ensembles in which every representative universe has exactly $N$ GW sources. However, since we do not know the correct value for $N$, we can construct a more realistic ensemble by combining ensembles with different values of $N$. In this combination, we assume that each source has a fixed mean-squared strain $h_s^2$.

A reasonable approach is to model the possible values of $N$ as a Poisson process with a mean number of sources given by some fixed $M > 0$. Thus, the ensemble contains representative universes with different values of $N$. For this Poisson model, the fraction of universes with $N$ GW sources is

$$f(N) = \frac{M^N}{N!} e^{-M}. \quad (5.1)$$

It is easy to check that this fraction is correctly normalized, since

$$\sum_{N=0}^{\infty} f(N) = 1. \quad (5.2)$$

The mean number of sources is

$$\langle N \rangle_N \equiv \sum_{N=0}^{\infty} N f(N) = M. \quad (5.3)$$

This is the justification for interpreting $M$ as the mean number of GW sources in the ensemble. Also useful is the second moment

$$\langle N^2 \rangle_N \equiv \sum_{N=0}^{\infty} N^2 f(N) = M(M+1). \quad (5.4)$$

Together, the first and second moments (5.3) and (5.4) imply that the variance in the expected number of sources is $\langle N^2 \rangle_N - \langle N \rangle_N^2 = M$. These are standard well-known results.

### A. Extending the discrete source ensemble to a Poisson process

One way to account for the variation in $N$ is to extend the discrete source calculation of Sec. IV. For this, we use the subscript $N$ to denote averages over the Poisson distribution of $N$, as already used in (5.3) and (5.4). The full ensemble average is

$$\langle Q \rangle = \langle Q \rangle_{\mathcal{A},\Omega,N} \equiv \sum_{N=0}^{\infty} f(N) \langle Q(N) \rangle_{\mathcal{A},\Omega}, \quad (5.5)$$

where $\langle Q(N) \rangle_{\mathcal{A},\Omega}$ denotes an ensemble average as calculated in the previous section, for an ensemble of universes containing $N$ sources.

Letting $h_s^2$ denote the mean-squared strain of a single GW source, the mean of the pulsar-averaged Hellings and Downs correlation is obtained by averaging the first moment of $\Gamma(\gamma)$ given in (4.10). We obtain

$$\langle \Gamma(\gamma) \rangle = \langle N h_s^2 \mu_u(\gamma) \rangle_N = \langle N \rangle_N h_s^2 \mu_u(\gamma) = M h_s^2 \mu_u(\gamma). \quad (5.6)$$

Thus, the mean-squared GW strain is the product of the expected number of sources $M$ and the expected squared GW strain $h_s^2$ of a single source.

The ensemble average of the pulsar-averaged correlation is obtained directly from (4.19)

$$\langle \Gamma(\gamma)\Gamma(\gamma')\rangle \equiv \langle \Gamma(\gamma)\Gamma(\gamma')\rangle_{\mathcal{A},\Omega,N}$$
$$= \langle N^2\rangle_N h_s^4 \mu_u(\gamma)\mu_u(\gamma') + \langle N\rangle_N \hbar_s^4 \mu_u(\gamma)\mu_u(\gamma') + 2(\langle N^2\rangle_N - \langle N\rangle_N)\hbar_s^4 \tilde{\mu}^2(\gamma,\gamma')$$
$$= M(M+1)h_s^4 \mu_u(\gamma)\mu_u(\gamma') + M\hbar_s^4 \mu_u(\gamma)\mu_u(\gamma') + 2M^2 \hbar_s^4 \tilde{\mu}^2(\gamma,\gamma'). \quad (5.7)$$

Subtracting the "square" of (5.6) gives the cosmic covariance

$$\sigma_{\cos}^2(\gamma,\gamma') = \langle \Gamma(\gamma)\Gamma(\gamma')\rangle - \langle \Gamma(\gamma)\rangle\langle \Gamma(\gamma')\rangle$$
$$= M(h_s^4 + \hbar_s^4)\mu_u(\gamma)\mu_u(\gamma')$$
$$+ 2M^2 \hbar_s^4 \tilde{\mu}^2(\gamma,\gamma'). \quad (5.8)$$

This is very similar to expression (4.23), which gives the cosmic covariance for exactly $N$ sources, if $N$ is replaced by the mean number $M$.

### B. Modifying the masking function

A second (equivalent) way to employ a Poisson distribution for the number of sources is via an ensemble of masking functions, as constructed in Sec. II. In analogy with (2.1), we define a masking function for $N$ sources by

$$\psi(\Omega) \equiv \frac{4\pi}{M}\sum_{j=1}^{N}\delta^2(\Omega,\Omega_j). \quad (5.9)$$

The normalization ensures that all GW sources make the same (mean-)squared strain contribution, regardless of the value of $N$. The overall factor proportional to $1/M$ is chosen, as we will show below, to ensure that the ensemble average of $\psi$ is unity.

The construction of the ensemble of masking functions proceeds as follows. (i) Initialize the ensemble to an empty set. (ii) Fix the mean number of sources $M > 0$. (iii) Select a random value of $N$ from the Poisson distribution with mean $M$ [so (5.1) is the probability that any particular value $N$ is obtained]. (iv) Select points $\Omega_1, \ldots, \Omega_N$ at random from a uniform distribution on the sphere. (v) Insert the masking function $\psi$ defined by (5.9) into the ensemble of masking functions. (vi) Return to step (iii).

To calculate ensemble averages involving $\psi$, we use the same methods and constructions as in Sec. II. Because (5.9) differs from (2.1) by a factor of $N/M$, the results may be obtained from those of Sec. II via a simple recipe: (A) scale each $\psi$ by a factor of $N/M$, then (B) compute the ensemble average over $N$ by means of (5.5).

Employing this recipe, the expected values of the spherical harmonic coefficients are obtained from (2.9) as follows:

$$\langle \psi_{lm}\rangle_N = \frac{\langle N\rangle_N}{M}\sqrt{4\pi}\,\delta_{l0}\delta_{m0} = \sqrt{4\pi}\,\delta_{l0}\delta_{m0}, \quad (5.10)$$

where the second equality follows from (5.3), The first moment (5.10) is the same as (2.9), implying that

$$\langle \psi(\Omega)\rangle_N = 1. \quad (5.11)$$

The second moments are found from (2.14) by using the recipe above:

$$\langle \psi_{lm}\psi_{l'm'}^*\rangle_N = \frac{4\pi}{M^2}\left[\langle N\rangle_N + \langle N^2 - N\rangle_N \delta_{l0}\delta_{m0}\right]\delta_{ll'}\delta_{mm'}$$
$$= 4\pi\left[\frac{1}{M} + \delta_{l0}\delta_{m0}\right]\delta_{ll'}\delta_{mm'}. \quad (5.12)$$

The first equality is obtained by multiplying the final expression in (2.14) with $N^2/M^2$ and averaging over the Poisson distribution. The second equality follows from application of (5.3) and (5.4).

The correlation function may be obtained from (5.12) or directly from the final expression in (2.15) by employing the recipe above. It is

$$\langle \psi(\Omega)\psi(\Omega')\rangle_N = \frac{4\pi}{M^2}\langle N\rangle_N \delta^2(\Omega,\Omega') + \langle N^2 - N\rangle_N/M^2$$
$$= \frac{4\pi}{M}\delta^2(\Omega,\Omega') + 1. \quad (5.13)$$

The covariance [22] [Eq. (9.3)] follows immediately from (5.11) and (5.13), and is

$$C(\Omega,\Omega') \equiv \langle \psi(\Omega)\psi(\Omega')\rangle_N - \langle \psi(\Omega)\rangle_N \langle \psi(\Omega')\rangle_N$$
$$= \frac{4\pi}{M}\delta^2(\Omega,\Omega'). \quad (5.14)$$

Hence, using the normalization conventions of [22] [Eq. (9.4)], the covariance may be expressed as a sum of Legendre polynomials

$$C(\Omega,\Omega') = \sum_{l=0}^{\infty}\left(\frac{2l+1}{4\pi}\right)C_l P_l(\Omega\cdot\Omega'), \quad (5.15)$$

with coefficients $C_l = 4\pi/M$ for all $l$. (The scale independence of these coefficients is why this is often described as a shot noise or "white" process.) This Poisson masking ensemble can be used to recover (5.8) via the same computation employed in (3.5).

## VI. ESTIMATING THE EFFECTS OF SOURCE DISCRETENESS AND GALAXY CLUSTERING ON THE RECOVERY OF THE HELLINGS AND DOWNS CURVE

Here, we use the results of the previous sections to estimate the effects of source discreteness and galaxy clustering on the recovery of the pulsar-averaged Hellings and Downs correlation. In what follows, cosmological lengths and densities are set to correspond to a dimensionless Hubble parameter $h_{100} = 0.68$ in agreement with the "Planck" value $H_0 = 68 \text{ km s}^{-1} \text{ Mpc}^{-1}$ for the present-day Hubble expansion rate.

### A. Finite number of sources (shot noise)

We first consider the effects of shot noise, resulting from the discreteness of the GW sources. Simulations of the GW background produced by pairs of orbiting supermassive black holes in the centers of merging galaxies show that the signal in the PTA band is usually dominated by a handful of bright sources, of order 10 or less, see e.g., [17]. This is a tiny fraction of the $O(10^6)$ such systems that are contributing to the unresolved component of the GW background.

To assess the impact of the resulting shot noise, Fig. 2 compares the cosmic standard deviation of the standard Gaussian ensemble with that for $N = 1, 3, 5, 10,$ and 100 discrete GW sources, using (3.6) and (4.23), respectively. In the comparison, we set $h_s^2 = h^2/N$ and $\hbar_s^4 = \hbar^4/N^2$ so that both ensembles have the same mean-squared strain at Earth, and use $h^2 = 1$ and $\hbar^2 = 0.5622$, which is appropriate for a binary inspiral power spectrum for timing residual measurements (see Table 3 in [10]). The fractional increase in the standard deviation arising from GW source discreteness (shot noise) is of order $1/N$.

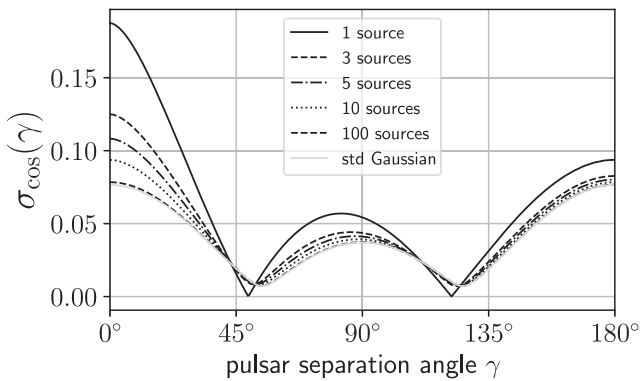

FIG. 2. Comparison of the cosmic standard deviation for the standard Gaussian ensemble with that for $N = 1, 3, 5, 10,$ and 100 discrete sources, calculated using (3.6) and (4.23), respectively. We have matched the squared strain at Earth for the standard Gaussian ensemble and discrete source models by setting $h_s^2 = h^2/N$ and $\hbar_s^4 = \hbar^4/N^2$. We have also set $h^2 = 1$ and $\hbar^2 = 0.5622$, which is appropriate for a binary inspiral power spectrum for timing residual measurements.

For $N = 1$, we have $\sigma_{\cos}(\gamma) = \hbar^2|\mu_u(\gamma)|$, with the zeroes of the Hellings and Downs curve noticeable in the plot. For $N \lesssim 10$ discrete sources, the difference between $\sigma_{\cos}(\gamma)$ for the discrete-source and standard Gaussian ensemble models is $\gtrsim 10\%$, which makes a noticeable difference to the cosmic variance, particularly at small angular separations.

### B. Galaxy clustering

The matter in the universe, and galaxies in particular, are not distributed as a random Poisson process. Galaxies tend to be grouped into clusters, and these clusters form patterns of filaments and voids, known as the cosmic web [21]. Filaments are threadlike formations where galaxies are concentrated, whereas voids are vast, relatively empty regions. This structure arises from the gravitational influence of dark matter and the universe's initial conditions, creating a complex large-scale galaxy distribution pattern.

The length scale of the clusters is typically $\approx 10$ Mpc, whereas the filaments and voids have $\approx 100$ Mpc length scales. Because the number density of normal galaxies is $n_g \approx 10^{-2}$ Mpc$^{-3}$ [32], these larger structures are formed from thousands or tens of thousands of galaxies. The pattern can be characterized by angular power spectra or by three-dimensional power spectra [19,20].

However, when the universe is observed with PTAs, the effects of this structure disappear. This is because the number density $n_G$ of massive galaxies relevant to PTAs (hosting black holes with masses greater than $10^9 M_\odot$) is observed/inferred to lie in the range $1.2 \times 10^{-4}$ Mpc$^{-3}$ [33] to $8 \times 10^{-4}$ Mpc$^{-3}$ [34]. This is shown as the lower band in Fig. 3. Because a typical binary spends only about 20 Myr orbiting in the PTA band, and the time between galactic mergers is Gyr, at most 1% of these galaxies would host PTA sources: a careful estimate yields 0.1% [35]. Thus, the number density $n_{\text{PTA}}$ of sources in the PTA band should lie in the range $10^{-7}$ Mpc$^{-3}$ to $10^{-5}$ Mpc$^{-3}$. This is shown as the upper band in Fig. 3, and is broadly consistent with the current PTA observational evidence [36].

This means that, on the average, these PTA sources (and host galaxies) are so far apart that the effects of the clustering and structure are no longer apparent: all that remains is the effect of the pointlike nature of sources, as previously estimated. Thus, the effects of galaxy clustering and the cosmic web are overwhelmed by the "shot noise" associated with the Poisson random process, exactly as we have described and calculated earlier in this paper. The amplitude of this shot noise is inversely proportional to the spatial number density of sources.

One way to see that shot noise dominates the effects of clustering for PTAs is via the present-day linear-theory matter power spectrum. This can be inferred from a variety of different cosmological probes [37] (Fig. 19), and is shown in Fig. 3. The solid line shows a smooth fit to the power spectrum of density fluctuations, as a function of spatial frequency (wave number). In comparison, the dashed horizontal lines show the levels of shot noise for massive galaxies that could host PTA sources (lower line, number density $n_G = 3 \times 10^{-4}$ Mpc$^{-3}$) and for the small fraction of those galaxies expected to host in-band PTA sources (upper line, assumed number density $n_{\rm PTA} = 10^{-6}$ Mpc$^{-3}$). The entire reasonable range of values corresponds to levels of shot noise which are larger than the fluctuations resulting from the matter power spectrum. Thus, the effects of galaxy clustering and structure are hidden: the PTA sources are too few and too far apart to reflect that structure.

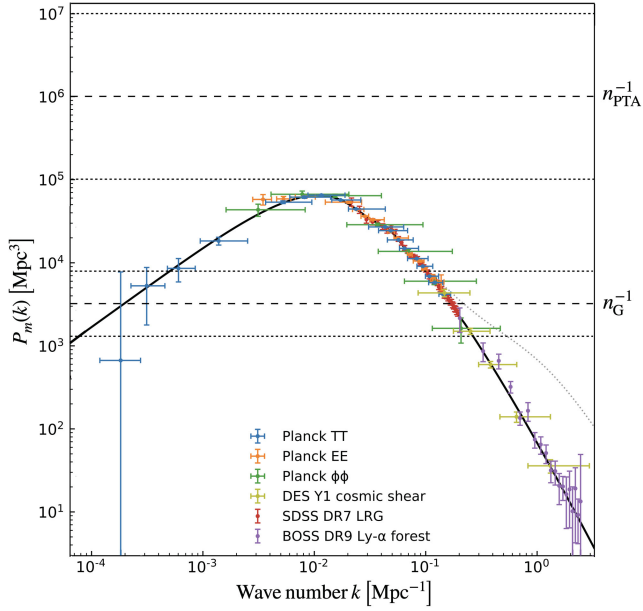

FIG. 3. The solid curve shows the linear-theory matter power spectrum (at zero redshift) inferred from different cosmological probes; the dotted curve shows the impact of nonlinear clustering at zero redshift. The lower horizontal dashed line shows the white "shot noise" power spectrum arising from the discreteness of galaxies that host massive black holes. The dotted interval around it corresponds to galaxy number densities $n_G$ in the range $1.2 \times 10^{-4}$ Mpc$^{-3}$ < $n_G$ < $8 \times 10^{-4}$ Mpc$^{-3}$, as discussed in the text. The upper horizontal dashed line and range are the quantities relevant for PTAs. These show the white power spectra of shot noise for the much smaller fraction of galaxies that host massive PTA binary sources. The horizontal dashed line is for a number density $n_{\rm PTA} = 10^{-6}$ Mpc$^{-3}$, and the dotted interval around it corresponds to source number densities in the range $10^{-5}$ Mpc$^{-3}$ < $n_{\rm PTA}$ < $10^{-7}$ Mpc$^{-3}$. Anywhere within this reasonable range, PTA source shot noise dominates clustering at all scales. (Plot adapted from [37] (Fig. 19), taking $h_{100} = 0.68$.)

The same conclusion was reached in [28] for current GW detectors such as LIGO and Virgo, but may not hold for next-generation terrestrial instruments such as Cosmic Explorer [38] or the Einstein Telescope [39]. For such terrestrial detectors, which operate in the audio-frequency band, the relevant GW sources are binary neutron stars (BNS), whose cosmological rate (per unit volume per unit time) is much larger than that of binary black holes (BBH) [40]. For current detectors [28], the shot noise dominates, hiding the underlying matter power spectrum. However, for next-generation detectors [38,39], with a decade of observation, the shot noise should lie near the peak of the power spectrum. This can be shown by simple order of magnitude estimates, as follows. The number density of normal galaxies is $n_g \approx 10^{-2}$ Mpc$^{-3}$, and in a decade we expect about one in a thousand of these to host a BNS merger. Thus, since future detectors are expected to be sensitive enough to observe all BNS mergers out to the Hubble distance, the level of shot noise is about $1/10^{-3} n_g \approx 10^5$ Mpc$^3$. Inspection of Fig. 3 shows that this is comparable to the value at the peak of the power spectrum. In such cases, where the shot noise has an amplitude whose magnitude is comparable to that of the power spectrum, it may be possible [27,29] to extract additional information [41] about the underlying power spectrum.

Note that Fig. 3 shows the power spectrum of linear-theory matter density perturbations; the nonlinear collapse into galaxies produces a bias that shifts the spectrum (solid curve) upwards. Typical estimates of the bias factor [42] are slightly less than 2. The upwards shift is proportional to the square of the bias factor, so these effects move the curve upwards by less than an order of magnitude. Thus, for PTAs, the shot noise would always dominate.

This conclusion can be verified by using the angular correlation function of galaxies. For this purpose, we derive an upper bound on the contribution of galactic clustering to the cosmic variance, employing a simple model for the angular power spectrum,

$$C_L = \begin{cases} C_0 & \text{for } L \leq L_{\max} \\ 0 & \text{for } L > L_{\max}. \end{cases} \quad (6.1)$$

To obtain an upper bound, start with the expression for the cosmic variance in terms of the angular power spectrum, see (9.10) and (9.4) in [22]:

$$\sigma_{\cos}^2(\gamma) = 2\hbar^4 \tilde{\mu}^2(\gamma) + \frac{C_0}{4\pi} h^4 \mu_u^2(\gamma)$$
$$+ \frac{1}{2} \hbar^4 \int_0^\pi d(\cos\beta) \sum_{L=0}^\infty \left(\frac{2L+1}{4\pi}\right)$$
$$\times C_L P_L(\cos\beta) \left[\mu^2(\gamma,\beta) + \mu^2(\bar\gamma,\bar\beta)\right]. \quad (6.2)$$

Begin by assuming that $C_L = C_0$ for *all* $L$, so that $L_{\max} = \infty$. Then the summation in (6.2) is easy to do since

$$\sum_{L=0}^\infty \left(\frac{2L+1}{2}\right) P_L(\cos\beta) = \delta(\cos\beta - 1), \quad (6.3)$$

which can be obtained from Eq. (4.2) of [22] by setting $x = \cos\beta$ and $x' = 1$. The integral over $\cos\beta$ is trivial, so the third term on the right-hand side of (6.2) simplifies to

$$\text{third term} = \frac{1}{4\pi}\hbar^4 C_0 \mu_u^2(\gamma), \quad (6.4)$$

where we used $\mu^2(\gamma, 0) = \mu_u^2(\gamma)$ and $\mu^2(\bar{\gamma}, \pi) = 0$.

Expression (6.4) for the third term of (6.2) implies an upper bound on the cosmic variance for the $C_L$ given in (6.1). This is because any single $C_L > 0$ makes a contribution to the cosmic variance which is necessarily nonnegative. A proof by contradiction follows. Suppose that for some $L$ value

$$\int_0^\pi d(\cos\beta)\left(\frac{2L+1}{4\pi}\right)P_L(\cos\beta)\left[\mu^2(\gamma,\beta) + \mu^2(\bar{\gamma},\bar{\beta})\right] < 0. \quad (6.5)$$

Then, by choosing that $C_L$ to be large enough to dominate the first two terms in (6.2), we could make $\sigma_{\cos}^2(\gamma) < 0$. This is a contradiction, since by definition $\sigma^2 \geq 0$.

So, we have proven that for the $C_L$ model given in (6.1)

$$\sigma_{\cos}^2(\gamma) \leq 2\hbar^4 \tilde{\mu}^2(\gamma) + \frac{1}{4\pi}(h^4 + \hbar^4)C_0\mu_u^2(\gamma). \quad (6.6)$$

If we now use

$$C_0 = 0.002, \qquad L_{\max} = 50, \quad (6.7)$$

to describe the effects of galaxy clustering [20], the contribution of these $C_L$'s to the cosmic variance is less than one percent of that for the standard Gaussian ensemble. Thus, galaxy clustering does not make a significant contribution to the cosmic variance of the pulsar-averaged Hellings and Downs correlation. This is consistent with the results of [23] (see Fig. 2 there) for these small values of $C_0$.

## VII. CONCLUSION

This paper uses the general method developed in [22,23] to assess the effects of (a) source discreteness and (b) galaxy clustering on the Hellings and Downs correlation. The impact is measured by comparing the cosmic variance of the pulsar-averaged correlation $\Gamma$ in the standard Gaussian ensemble to the same quantity in an ensemble which models (a) or (b). Here, fluctuations in $\Gamma$ quantify "how closely" the Hellings and Downs correlation is expected to approach the Hellings and Downs curve. This provides an order-of-magnitude estimate of the impact of these effects.

For (a) we compared ensembles containing a Poisson distribution of $N$ discrete sources to the Gaussian ensemble. The latter corresponds to a very large number of very weak sources. We found (see Fig. 2) that for typical numbers of strong PTA sources $N \approx 10$ this shot noise effect increases the cosmic standard deviation of the pulsar-averaged Hellings and Downs correlation by an amount $1/N \approx 10\%$. For (b) we compared the Gaussian ensemble without source correlations to an ensemble with typical galactic structure correlations. Here, the effects are very small, less than 1%, and completely dominated by the shot noise contributions (a) [see Fig. 3 and text following (6.7)]. These results provide a useful quantitative test and point of comparison for more authentic computer simulations.

Our model (ensemble) of $N$ discrete GW sources assumes that these have a Gaussian distribution of amplitudes, drawn from a power spectrum. Physically, this corresponds to picking $N$ lines of sight (directions on the sphere). Then, for each of these sky directions, a large number of independently radiating GW sources are stacked up "on top of each other" along that line of sight. For each direction, enough sources are stacked that the central limit theorem applies, and the resulting GW amplitudes along each of the $N$ directions has the same Gaussian distribution.

These calculations could be improved by using a more realistic ensemble, i.e., by modifying the statistical ensembles to make them more realistic. For this, along each of the $N$ directions, there would be one GW source, which radiates at a single frequency with a specific (but unknown) phase and polarization. Rather than selecting the amplitudes from the same distributions, one could use distributions whose mean-squared amplitudes are largest for the "closest" source and are smaller for more distant ones. Such calculations are not difficult to carry out, and while we do not expect that this will change the order of magnitudes of the different effects, it would be helpful to quantify this.

*Note added in proof.*—Recent work by two of the authors shows that the "zero lag" Hellings and Downs estimator based on the uniform time averages of $Z_p Z_q$ [as defined before (4.4)] may have its variance reduced, and thus be improved, by suitable frequency-space weighting [44]. This does not affect our conclusions.

## ACKNOWLEDGMENTS

B. A. would like to thank Eichiro Komatsu for helpful discussions regarding the relative contributions of shot noise versus clustering, Luke Zoltan Kelley for a selective review of the literature on the expected number densities of PTA host galaxies and sources, Chiara Mingarelli for guidance on the same topic, Sumit Kumar for helpful discussions and correspondence regarding the observability of the matter power spectrum with current and future terrestrial GW detectors, and the anonymous referee for several helpful suggestions. D. A. acknowledges financial support from the Actions de Recherche Concertées (ARC) and Le Fonds spécial pour la recherche (FSR) of the Féderation Wallonie-Bruxelles. J. D. R. acknowledges financial support from the NSF Physics Frontier Center Award PFC-2020265 and startup funds from the University of Texas Rio Grande Valley.

SOURCE ANISOTROPIES AND PULSAR TIMING ARRAYS PHYS. REV. D **110,** 123507 (2024)